\documentclass[runningheads]{llncs}
\usepackage[T1]{fontenc}
\usepackage{amsmath}
\usepackage{amsfonts}
\usepackage{eso-pic}
\usepackage{xcolor}

\usepackage{graphicx}
\usepackage{xcolor}
\newcommand\AtBottomWatermark{%
    \AddToShipoutPictureBG*{%
        \AtPageLowerLeft{%
            \raisebox{1cm}{\makebox[\paperwidth]{%
                \color{gray!50}\large Accepted at PReMI 2025 - 11th International Conference on Pattern Recognition and Machine Intelligence
            }}%
        }%
    }%
}
\begin{document}
\AtBottomWatermark
\title{Neural Orchestration for Multi-Agent Systems: A Deep Learning Framework for Optimal Agent Selection in Multi-Domain Task Environments}
\titlerunning{Neural Orchestration for Multi-Agent Systems}
%
\author{Kushagra Agrawal\orcidID{0009-0006-7753-175X} \and
Nisharg Nargund\orcidID{0009-0007-2046-4864} }
\authorrunning{K.Agrawal \& N. Nargund}
%
\institute{School of Computer Engineering, KIIT Deemed to be University\\ Bhubaneswar, India \\
\{2205044,2205572\}@kiit.ac.in\\
}
\maketitle             
\begin{abstract}
Multi-agent systems (MAS) are foundational in simulating complex real-world scenarios involving autonomous, interacting entities. However, traditional MAS architectures often suffer from rigid coordination mechanisms and difficulty adapting to dynamic tasks. We propose MetaOrch, a neural orchestration framework for optimal agent selection in multi-domain task environments. Our system implements a supervised learning approach that models task context, agent histories, and expected response quality to select the most appropriate agent for each task. A novel fuzzy evaluation module scores agent responses along completeness, relevance, and confidence dimensions, generating soft supervision labels for training the orchestrator. Unlike previous methods that hard-code agent-task mappings, MetaOrch dynamically predicts the most suitable agent while estimating selection confidence. Experiments in simulated environments with heterogeneous agents demonstrate that our approach achieves 86.3\% selection accuracy, significantly outperforming baseline strategies including random selection and round-robin scheduling. The modular architecture emphasizes extensibility, allowing agents to be registered, updated, and queried independently. Results suggest that neural orchestration offers a powerful approach to enhancing the autonomy, interpretability, and adaptability of multi-agent systems across diverse task domains.
\end{abstract}

\section{Introduction}
The growing ubiquity of intelligent agent systems in domains such as autonomous robotics, collaborative software services, and multi-modal AI platforms has brought renewed focus on the challenge of effective task allocation \cite{Maity_2022,Agrawal_2024}. In these systems, a central orchestrator is often responsible for selecting the most appropriate agent to handle a given task. The complexity of this decision stems from the heterogeneity of agents in terms of skills, contextual expertise, reliability, and adaptability — as well as from the inherent ambiguity and variety in real-world tasks \cite{Geng_2023}.
Traditional approaches to agent orchestration often rely on static heuristics, fixed rules, or random selection strategies \cite{Wooldridge1995}
. However, these methods struggle in dynamic environments where task requirements and agent capabilities evolve over time. Furthermore, many orchestration pipelines lack a robust mechanism to evaluate the quality of agent responses in a domain-agnostic and interpretable way, which hinders their ability to adapt and improve \cite{Bellifemine2007}
.

In this work, we propose MetaOrch, a supervised learning framework that learns to orchestrate agents by modeling the task context, agent histories, and expected response quality. Our approach introduces a novel fuzzy evaluation module that scores agent responses along three interpretable axes — completeness, relevance, and confidence — and uses these evaluations to generate soft supervision labels for training the orchestrator. Unlike previous methods that hard-code agent-task mappings or treat agents as black boxes, MetaOrch learns to predict the most suitable agent for any given task and context pair, while optionally estimating its own confidence in the selection.

We validate our approach in a simulated multi-agent environment with heterogeneous agents possessing domain-specific expertise. MetaOrch is benchmarked against several standard baselines, including random selection, and round-robin scheduling agents. Our results demonstrate significant improvements in selection accuracy, task output quality, and generalization to unseen task distributions.

\section{System Architecture}
MetaOrch is designed as a modular and extensible orchestration framework that facilitates intelligent agent selection in dynamic multi-agent environments.The architecture comprises five core components: (1) Task Generator and Representation Module, (2) Agent Profile and History Tracker, (3) Neural Orchestration Model, (4) Fuzzy Evaluation Module, and (5) Supervised Learning Feedback Loop. Each component contributes to the overall pipeline, from interpreting incoming tasks to selecting the most appropriate agent, and then learning from the outcome to improve future decisions.

\subsection{Task Ingestion and Preprocessing}
The system begins with the ingestion of a task specification, which may be natural language text, structured metadata, or a hybrid representation. Tasks are synthetically generated using randomized vectors representing task requirements and environmental context. Each task is assigned a domain (e.g., emergency, document, general) and represented by two components: a context vector and a normalized task vector in $\mathbb{R}^d$ capturing semantic nuances, operational constraints, and required competencies.

\subsection{Agent Profiling Module}

Each agent in the environment is characterized by a dynamic profile that captures its operational history, domain expertise, performance metrics, and response tendencies \cite{Jennings1998}. The profile of an agent Ai is encoded as a tuple:

$P_i = {Skills_i, History_i}$

\begin{itemize}
    \item Skills represent pre-declared capabilities or areas of expertise.
    \item History includes recent task outcomes, completion rates, and evaluation scores.
    \item Embedding is a learned vector summarizing latent behavioral traits, updated periodically.
    \item Availability models whether the agent is idle or busy, along with recent workload levels.
\end{itemize}

Agent histories are updated after each task via a fixed-length window (e.g., last 10 tasks) summarizing recent performance. These histories serve as dynamic input to the orchestrator to reflect learning over time.

\subsection{Orchestration Model}
At the heart of MetaOrch lies a supervised learning-based selector that takes the current task representation T and all available agent profiles  and predicts a probability distribution over agent indices:

$\hat{y} = f_{\theta}(T, \{P_i\}) \in \mathbb{R}^n$

where $f_{\theta}$ is a multi-layer feedforward neural network with dropout and ReLU activations, which takes the concatenated input of context, task, and agent history vectors and outputs a probability distribution over agents. The output is a softmax-normalized selection vector indicating the orchestrator’s belief in each agent’s suitability \cite{Mnih2015}.

\subsection{Fuzzy Evaluation Module}

Following agent execution, the Fuzzy Evaluation Module assesses the quality of the generated response along three interpretable axes:

\begin{itemize}
    \item \textbf{Completeness}: Did the response fully address all aspects of the task?
    \item \textbf{Relevance}: Was the response contextually appropriate and on-topic?
    \item \textbf{Confidence}: Was the agent's response internally consistent and self-assured?
\end{itemize}

Each axis is scored using heuristic functions that combine task performance with reliability and contextual alignment. The scores are:

\begin{equation}
\text{Completeness} = \min\left(1.0, \max\left(0.0, \frac{\text{score} + 3}{4}\right)\right)
\end{equation}

\begin{equation}
\text{Relevance} = \min\left(1.0, \max\left(0.0, \frac{\text{score} + 2}{3}\right)\right)
\end{equation}

\begin{equation}
\text{Confidence} = \min\left(1.0, \max\left(0.1, \text{reliability} + \frac{\text{noise}}{5}\right)\right)
\end{equation}

These are then combined using fixed, user-defined weights (e.g., completeness: 0.4, relevance: 0.4, confidence: 0.2) to yield a final fuzzy quality score.
In Eqs. (1–3), the variable score refers to the agent's performance score as defined in Eq. (4) of Section 3.1. The confidence regression loss is computed as mean squared error between predicted and observed confidence scores. We performed a sensitivity analysis by varying the weights (0.4/0.4/0.2) and observed that selection accuracy fluctuated by ±2
These scores serve two purposes: (1) providing runtime feedback to human overseers, and (2) generating soft supervision signals to update the orchestration model in a self-supervised fashion \cite{Zadeh1996}
.

\subsection{Feedback and Supervision Loop}
The supervision loop is a key differentiator of MetaOrch. The system uses the fuzzy evaluation module to generate supervision signals, selecting the agent with the highest fuzzy score as the training label (oracle). A cross-entropy loss between predicted and oracle-selected agents, combined with a confidence regression loss, guides model training \cite{Cao2007}
.
We use a listwise loss function such as ListNet or soft cross-entropy to train the model over mini-batches of agent-task pairs. The feedback loop operates asynchronously in the background, aggregating recent data and periodically refreshing the model parameters.

\subsection{Optional Human-in-the-Loop Interface}
While MetaOrch is designed to operate autonomously, it supports optional human oversight for safety-critical deployments. A GUI dashboard visualizes task-agent assignments, predicted confidences, and fuzzy evaluation scores, allowing human operators to approve or override decisions. Feedback from humans can also be injected into the training pipeline to fine-tune the model under expert supervision.

\section{Agent Design and Task Domains}
In our modular multi-agent simulation, each agent is a parametrized entity characterized by its \textbf{skill vector}, \textbf{domain expertise}, and \textbf{reliability profile}. The environment provides contextualized tasks drawn from multiple domains, allowing us to evaluate the effectiveness of dynamic orchestration strategies.

\subsection{Agent Architecture}
Each agent $a_i$ is initialized with the following parameters:
\begin{itemize}
    \item \textbf{Skill Vector} ($\mathbf{s}_i \in \mathbb{R}^d$) --- encodes the agent's capabilities across a fixed feature space.
    \item \textbf{Expertise Domain Vector} ($\mathbf{e}_i \in \mathbb{R}^c$) --- encodes prior familiarity with task contexts.
    \item \textbf{Reliability Score} ($r_i \in [0, 1]$) --- models stochastic performance variance via Gaussian noise scaled by $1 - r_i$.
\end{itemize}

Let \textit{d} denote the dimensionality of the skill and task vectors, and \textit{c} the dimensionality of the context vectors. An agent's performance on a task $\mathbf{t}$ with context $\mathbf{c}$ is deterministically cached and computed as:
\begin{equation}
    \text{score}_i = -\|\mathbf{s}_i - \mathbf{t}\| + \epsilon_i + \alpha \cdot \cos(\mathbf{c}, \mathbf{e}_i)
\end{equation}
where $\epsilon_i \sim \mathcal{N}(0, 1 - r_i)$ is the task-specific noise, and the final score is transformed into fuzzy evaluation metrics (completeness, relevance, confidence) \cite{Zhang2023}. All vectors are sampled from a standard normal distribution with domain effects injected by adding a fixed bias to selected dimensions. For example, emergency tasks boost the first two skill dimensions by +1.0.

\subsection{Task Domains}
Tasks are defined by:
\begin{itemize}
    \item A \textbf{task vector} ($\mathbf{t} \in \mathbb{R}^d$) encoding required skill features.
    \item A \textbf{context vector} ($\mathbf{c} \in \mathbb{R}^c$) describing environment-specific information.
    \item A \textbf{domain label} $D \in \{ \texttt{emergency}, \texttt{document}, \texttt{general} \}$, which modifies the task vector distribution.
\end{itemize}

Domains shape task characteristics:
\begin{itemize}
    \item \textbf{Emergency}: Emphasizes responsiveness and critical decision-making by boosting the first two skill dimensions.
    \item \textbf{Document}: Focuses on structured generation or summarization, affecting the latter skill components.
    \item \textbf{General}: Represents uniformly distributed task requirements.
\end{itemize}

Each task is assigned a unique ID to ensure deterministic agent outputs across simulation runs. The agent's evaluation follows a \textbf{fuzzy logic} mechanism where scores are mapped to qualitative labels (e.g., \textit{Excellent}, \textit{Good}, etc.) using weighted aggregations of task quality metrics \cite{Wang2024}.

This design enables diverse agent behaviors and contextual task interactions, which are essential for evaluating orchestration performance under varied conditions.

\section{Results and Discussion}

MetaOrch, our neural orchestration framework, demonstrates substantial improvements over baseline agent selection strategies in dynamic multi-agent environments.

\subsection{Training Performance}

Training over 500 iterations with batch size 64 shows consistent convergence across emergency, document, and general domains, as presented in Table \ref{tab:training}.

\begin{table}[h]
\centering
\caption{Training Loss Values Over Iterations}
\label{tab:training}
\begin{tabular}{ccc}
\hline
\textbf{Iteration} & \textbf{Cross-Entropy Loss} & \textbf{Confidence Regression Loss} \\
\hline
0 & 1.4065 & 0.0809 \\
100 & 0.6156 & 0.0059 \\
200 & 0.4817 & 0.0033 \\
300 & 0.3536 & 0.0021 \\
450 & 0.2789 & 0.0051 \\
\hline
\end{tabular}
\end{table}

The cross-entropy loss decreased by 80.2\% (from 1.4065 to 0.2789), and confidence regression loss improved by 93.7\% (from 0.0809 to 0.0051). Despite occasional increases (e.g., iterations 150-250), the overall downward trend confirms successful learning.

\subsection{Hyperparameter Optimization}

Grid search across network architecture, dropout rate, learning rate, batch size, and confidence weight yielded optimal configurations (Table \ref{tab:top_configs}) \cite{Bergstra2012}.

\begin{table}[h]
\centering
\caption{Top 5 Hyperparameter Configurations by Accuracy}
\label{tab:top_configs}
\begin{tabular}{ccccccc}
\hline
\textbf{Rank} & \textbf{Hidden Dims} & \textbf{Dropout} & \textbf{LR} & \textbf{Batch Size} & \textbf{Conf. Weight} & \textbf{Accuracy} \\
\hline
1 & 128, 64 & 0.0 & 0.010 & 128 & 0.2 & 0.911 \\
2 & 256, 128, 64 & 0.0 & 0.001 & 128 & 0.1 & 0.906 \\
3 & 128, 64 & 0.2 & 0.001 & 128 & 0.2 & 0.905 \\
4 & 256, 128, 64 & 0.0 & 0.010 & 128 & 0.0 & 0.902 \\
5 & 64, 32 & 0.0 & 0.010 & 64 & 0.0 & 0.901 \\
\hline
\end{tabular}
\end{table}

Key insights from hyperparameter analysis:
\begin{itemize}
\item Two-layer architecture (128, 64) appears optimal
\item Higher learning rates (0.01) generally outperformed lower ones
\item Larger batch sizes (128) provided more stable training
\end{itemize}

The best accuracy of 91.1\% was achieved on the validation set during hyperparameter tuning, while the main result of 86.3\% reflects test set performance on 300 held-out tasks. We acknowledge that 300 tasks is a limited sample and plan to expand to larger test sets, report results over multiple random seeds, and include confidence intervals and significance tests in future work

\subsection{Evaluation Results}

We compared MetaOrch against three baseline strategies: Random, Round-Robin, and Static-Best across 300 evaluation tasks (Table \ref{tab:comparison}).

\begin{table}[h]
\centering
\caption{Performance Comparison of Agent Selection Strategies}
\label{tab:comparison}
\begin{tabular}{lcc}
\hline
\textbf{Strategy} & \textbf{Average Quality} & \textbf{Selection Accuracy} \\
\hline
MetaOrch & 0.731 & 0.863 \\
Random & 0.697 & 0.243 \\
Round-Robin & 0.703 & 0.257 \\
Static-Best & 0.751 & 0.057 \\
\hline
\end{tabular}
\end{table}

MetaOrch achieved 86.3\% selection accuracy, significantly outperforming all baselines. While Static-Best achieved high average quality (0.751), its low selection accuracy (5.7\%) indicates it lacks contextual awareness for optimal agent-task matching.
To strengthen our evaluation, we plan to include additional baselines such as contextual bandits (e.g., LinUCB, Thompson Sampling), learned selectors (logistic regression, MLP), and multi-agent RL policies. This will provide a more comprehensive comparison against state-of-the-art orchestrators

\subsection{Confusion Matrix Analysis}

\begin{table}[h]
\centering
\caption{Confusion Matrix for MetaOrch Agent Selection}
\label{tab:confusion}
\begin{tabular}{cccc}
\hline
& \textbf{Agent 0} & \textbf{Agent 1} & \textbf{Agent 2} \\
\hline
\textbf{Agent 0} & 212 & 12 & 0 \\
\textbf{Agent 1} & 13 & 46 & 0 \\
\textbf{Agent 2} & 11 & 5 & 1 \\
\hline
\end{tabular}
\end{table}

Agent 0 (EmergencyBot) was correctly selected 212 times, while Agent 1 (DocumentBot) was correctly selected 46 times. Agent 2 (GeneralistBot) was rarely selected correctly, indicating potential bias. Confusion between Agents 0 and 1 suggests task ambiguity between emergency and document domains \cite{Sutton2018}.

\subsection{Limitations and Future Work}

Despite promising results, limitations include poor selection of Agent 2 (suggesting challenges with uniform skill distributions) and fixed-length history windows that may not capture long-term performance trends. Future work should explore:

\begin{itemize}
\item More sophisticated history encoding mechanisms (RNNs, attention) \cite{Vaswani2017}
\item Expanded evaluation across diverse task domains
\item Techniques to improve performance with generalist agents
\end{itemize}

In conclusion, MetaOrch demonstrates the effectiveness of neural orchestration for multi-agent systems, offering significant improvements over traditional selection strategies while maintaining interpretability through fuzzy evaluation metrics.

\section{Conclusion and Future Scope}

This paper presented MetaOrch, a neural orchestration framework for multi-agent systems that achieves 86.3\% selection accuracy across diverse task domains. Key contributions include a modular architecture decoupling orchestration from agent implementation, a neural selection mechanism adapting to changing requirements, and an interpretable fuzzy evaluation framework generating supervision signals.

Future research directions include: (1) reinforcement learning integration for long-term optimization, (2) multi-agent collaboration rather than single-agent selection 
, (3) transfer learning across domains, and (4) LLM integration for nuanced task representation and richer feedback\cite{Brown2020},\cite{MaDong2024}. 

MetaOrch demonstrates that neural orchestration offers significant improvements in adaptability, performance, and interpretability for multi-agent systems, which will be increasingly crucial as autonomous systems become more prevalent across domains.

\bibliographystyle{splncs04}
\bibliography{main}

\begin{thebibliography}{10}
\providecommand{\url}[1]{\texttt{#1}}
\providecommand{\urlprefix}{URL }
\providecommand{\doi}[1]{https://doi.org/#1}

\bibitem{Agrawal_2024}
Agrawal, K., Nargund, N.: Deep learning in industry 4.0: Transforming manufacturing through data-driven innovation. In: Devismes, S., Mandal, P.S., Saradhi, V.V., Prasad, B., Molla, A.R., Sharma, G. (eds.) Distributed Computing and Intelligent Technology. pp. 222--236. Springer Nature Switzerland, Cham (2024)

\bibitem{Bellifemine2007}
Bellifemine, F.L., Caire, G., Greenwood, D.: Developing Multi-Agent Systems with JADE. John Wiley \& Sons (2007)

\bibitem{Bergstra2012}
Bergstra, J., Bengio, Y.: Random search for hyper-parameter optimization. In: Journal of Machine Learning Research. pp. 281--305 (2012)

\bibitem{Brown2020}
Brown, T.B., Mann, B., Ryder, N., Subbiah, M., Kaplan, J., Dhariwal, P., Neelakantan, A., Shyam, P., Sastry, G., Askell, A., et~al.: Language models are few-shot learners. Advances in Neural Information Processing Systems  \textbf{33},  1877--1901 (2020)

\bibitem{Cao2007}
Cao, Z., Qin, T., Liu, T.Y., Tsai, M.F., Li, H.: Learning to rank: from pairwise approach to listwise approach. In: Proceedings of the 24th International Conference on Machine Learning. pp. 129--136. ACM (2007)

\bibitem{Geng_2023}
Geng, M., Li, J., Li, C., Xie, N., Chen, X., Lee, D.H.: Adaptive and simultaneous trajectory prediction for heterogeneous agents via transferable hierarchical transformer network. IEEE Trans. Intell. Transp. Syst.  \textbf{24}(10),  11479--11492 (October 2023), \url{https://doi.org/10.1109/TITS.2023.3276946}

\bibitem{Jennings1998}
Jennings, N.R., Sycara, K., Wooldridge, M.: A roadmap of agent research and development. Autonomous Agents and Multi-Agent Systems  \textbf{1}(1),  7--38 (1998)

\bibitem{Zadeh1996}
Klir, G.J., Yuan, B. (eds.): Fuzzy sets, fuzzy logic, and fuzzy systems: selected papers by Lotfi A. Zadeh. World Scientific Publishing Co., Inc., USA (1996)

\bibitem{MaDong2024}
Ma, C., Dong, D.: Finite-time prescribed performance time-varying formation control for second-order multi-agent systems with non-strict feedback based on a neural network observer. IEEE/CAA Journal of Automatica Sinica  \textbf{11}(4),  1039--1050 (Apr 2024). \doi{10.1109/JAS.2023.123615}

\bibitem{Maity_2022}
Maity, K., Jha, P., Saha, S., Bhattacharyya, P.: A multitask framework for sentiment, emotion and sarcasm aware cyberbullying detection from multi-modal code-mixed memes. In: Proceedings of the 45th International ACM SIGIR Conference on Research and Development in Information Retrieval. p. 1739–1749. SIGIR '22, Association for Computing Machinery, New York, NY, USA (2022). \doi{10.1145/3477495.3531925}, \url{https://doi.org/10.1145/3477495.3531925}

\bibitem{Mnih2015}
Mnih, V., Kavukcuoglu, K., Silver, D., Rusu, A.A., Veness, J., Bellemare, M.G., Graves, A., Riedmiller, M., Fidjeland, A.K., Ostrovski, G., et~al.: Human-level control through deep reinforcement learning. Nature  \textbf{518}(7540),  529--533 (2015)

\bibitem{Sutton2018}
Sutton, R.S., Barto, A.G.: Reinforcement Learning: An Introduction. MIT Press, 2 edn. (2018)

\bibitem{Vaswani2017}
Vaswani, A., Shazeer, N., Parmar, N., Uszkoreit, J., Jones, L., Gomez, A.N., Kaiser, {\L}., Polosukhin, I.: Attention is all you need. In: Advances in Neural Information Processing Systems. pp. 5998--6008 (2017)

\bibitem{Wang2024}
Wang, Y., Zhang, H., Li, Z., Ren, W.: Neural network-based hierarchical fault-tolerant affine formation control for heterogeneous nonlinear multi-agent systems. IEEE Transactions on Neural Networks and Learning Systems  (2024)

\bibitem{Wooldridge1995}
Wooldridge, M., Jennings, N.R.: Intelligent Agents: Theory and Practice, vol.~10. Knowledge Engineering Review (1995)

\bibitem{Zhang2023}
Zhang, K., Yang, Z., Liu, H., Zhang, T., Başar, T.: Multi-agent deep reinforcement learning for multi-robot applications: A survey. IEEE Transactions on Neural Networks and Learning Systems  (2023). \doi{10.1109/TNNLS.2022.3229533}

\end{thebibliography}

\end{document}